\documentclass[conference]{IEEEtran}
\IEEEoverridecommandlockouts
\usepackage{subcaption}
\usepackage{cite}
\usepackage{amsmath,amssymb,amsfonts}
\usepackage{algorithmic}
\usepackage{graphicx}
\usepackage{textcomp}
\usepackage{xcolor}
\def\BibTeX{{\rm B\kern-.05em{\sc i\kern-.025em b}\kern-.08em
    T\kern-.1667em\lower.7ex\hbox{E}\kern-.125emX}}
\begin{document}

\title{Enhancing Privacy and Security of Autonomous UAV Navigation\\}

\author{\IEEEauthorblockN{Vatsal Aggarwal$^{*}$}
\IEEEauthorblockA{
\textit{University at Buffalo}\\
New York, USA \\
vatsalag@buffalo.edu
}
\and
\IEEEauthorblockN{Arjun Ramesh Kaushik$^{*}$}
\IEEEauthorblockA{
\textit{University at Buffalo}\\
New York, USA \\
kaushik3@buffalo.edu
}
\and
\IEEEauthorblockN{Charanjit Jutla}
\IEEEauthorblockA{
\textit{IBM Research}\\
USA \\
csjutla@us.ibm.com
}
\and
\IEEEauthorblockN{Nalini Ratha}
\IEEEauthorblockA{
\textit{University at Buffalo}\\
New York, USA \\
nratha@buffalo.edu
}
\thanks{* First two authors contributed equally to this work.}
}

\maketitle

\begin{abstract}
Autonomous Unmanned Aerial Vehicles (UAVs) have become essential tools in defense, law enforcement, disaster response, and product delivery. These autonomous navigation systems require a wireless communication network, and of late are deep learning based. In critical scenarios such as border protection or disaster response, ensuring the secure navigation of autonomous UAVs is paramount. But, these autonomous UAVs are susceptible to adversarial attacks through the communication network or the deep learning models - eavesdropping / man-in-the-middle / membership inference / reconstruction. To address this susceptibility, we propose an innovative approach that combines Reinforcement Learning (RL) and Fully Homomorphic Encryption (FHE) for secure autonomous UAV navigation. This end-to-end secure framework is designed for real-time video feeds captured by UAV cameras and utilizes FHE to perform inference on encrypted input images. While FHE allows computations on encrypted data, certain computational operators are yet to be implemented. Convolutional neural networks, fully connected neural networks, activation functions and OpenAI Gym Library are meticulously adapted to the FHE domain to enable encrypted data processing. We demonstrate the efficacy of our proposed approach through extensive experimentation. Our proposed approach ensures security and privacy in autonomous UAV navigation with negligible loss in performance.


\end{abstract}


\section{Introduction}

Unmanned Aerial Vehicles (UAVs), commonly referred to as drones, are defined as aircrafts that operate without any human onboard. UAVs have brought about transformative changes across various industries, providing unmatched capabilities in surveillance, reconnaissance, disaster response, and product delivery \cite{drones6060147}. As the demand for more complex tasks performed by UAVs grows, so do the challenges in their development, particularly in striving for fully autonomous operation with minimal human intervention. Effective deployment of autonomous UAVs requires intricate path planning, obstacle detection, intelligent maneuverability and a wireless communication network. Research efforts on autonomous navigation in UAVs for visual mapping, obstacle detection, and path planning have gravitated towards deep neural networks \cite{doi:10.1080/10095020.2017.1420509}\cite{8993742}\cite{8600371}\cite{9729807}. In critical scenarios such as surveillance and disaster response, a secure wireless communication network to ensure secure navigation is imperative. In addition to susceptible wireless networks, deep neural networks in drones are also vulnerable to adversarial attacks \cite{makdad_survey_drone} \cite{guo_wang_uav_secure}.

Autonomous drones are exposed to various adversarial threats, such as - eavesdropping, traffic analysis, man-in-the-middle, and backdoor access \cite{types_of_drone_attacks}. From a deep learning perspective, attacks can be broadly classified into our types: membership inference, reconstruction, property inference, and model extraction \cite{survey_ml_attacks}. In our research, we specifically address the scenario where an attacker can intercept communication between the drone and its navigation server, posing a potential risk to the UAV's secure operation. Our primary focus is on establishing secure and private communication channels for autonomous drone navigation.


\begin{figure*}[htbp]
\centerline{\includegraphics[width=1\textwidth]{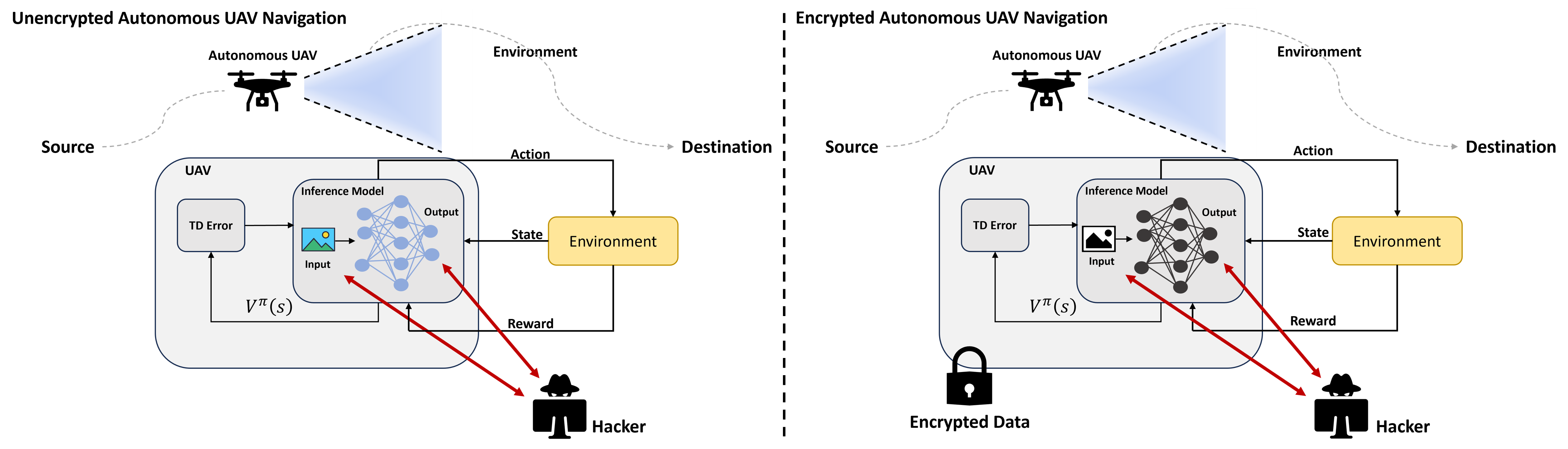}}
\caption{\textbf{Overview:} In an ordinary scenario the UAV is vulnerable to snooping attacks, as the attacker can directly steal the information. Or, query the model to infer target information, launching a model inversion attack. In our approach, the input is encrypted and the inference happens \textit{in the encrypted domain}. Hence, the attacker is unable to exploit any meaningful information from  the system.}
\label{intro_fig}
\end{figure*}

Previous works have explored computer vision-based autonomous UAV systems \cite{doi:10.1080/10095020.2017.1420509}, whereas, recent efforts take a Reinforcement Learning (RL) approach \cite{8993742} \cite{8600371} \cite{9729807}. In our work, we adopt the Actor-Critic model with Proximal Policy Optimization (PPO) as the policy gradient algorithm to demonstrate a solution addressing the privacy and security challenges in autonomous UAVs. We propose a novel end-to-end framework combining RL and Fully Homomorphic Encryption (FHE) to perform secure inferencing on homomorphically encrypted inputs. Using FHE in our proposed method ensures that the navigation result is only disclosed to authorized parties with the secret key for Homomorphic Encryption, and inputs themselves are secured through encryption during the navigation process. We show through our experiments that, with FHE, the navigation results are unaffected while guaranteeing utmost security. 

When implemented properly, FHE helps achieve a high level of security. However, its implementation of mathematical operations is limited. To this end, in this paper, we adapt the RL model to the encrypted domain to facilitate the processing of encrypted real-time images captured by UAV cameras (Fig. \ref{intro_fig}). Key aspects of our approach include transforming convolutional layers into spectral domain operations, utilizing generalized matrix multiplication in fully connected layers, and customizing activation functions as polynomial approximations/comparators. Since the RL framework utilizes OpenAI Gym Library to derive the navigational steps from the extracted image features, we adapt the Library to the encrypted domain as well. A simple multi-layer perceptron is trained to replicate the OpenAI Gym library and its weights are used during inferencing in the encrypted domain. Remarkably, our end-to-end secure framework shows a negligible loss in performance.


The rest of the paper is organized as follows - Section II covers the basics of FHE and Section III touches on the related work in this field. Section IV highlights how each component of the deep learning model is uniquely adapted for encrypted data handling. Sections V and VI detail the Mean Absolute Errors (MAEs) and time taken for each block in our novel framework. This work not only addresses immediate privacy and security concerns associated with UAVs but also lays the foundation for a new paradigm in autonomous aerial systems. By prioritizing privacy and security through FHE, our approach paves the way for deploying UAVs in sensitive domains where data confidentiality is paramount. 




\section{FHE basics}
Homomorphic Encryption (HE) is a cryptographic system that enables computations on encrypted data without the need for decryption. In this system, two key components are utilized: public key \(p_k\) and secret key \(s_k\). Encryption and decryption operations are denoted by \(E\) and \(D\), respectively. Consider plaintext values \(x\) and \(y\), and their corresponding ciphertexts, denoted as \({x'} = E(x, p_k)\) and \({y'} = E(y, p_k)\). HE empowers the computation of various operations directly on encrypted ciphertexts. For instance, the addition of encrypted values \(D(x' + y'\)) is approximately equivalent to the addition of the original plaintext values $x + y$. Likewise, the multiplication of encrypted values \(D(x' * y'\)) is approximately equivalent to $x * y$.

Among various HE schemes, Fully Homomorphic Encryption (FHE) can support computations on ciphertexts of any depth and complexity. This unique characteristic makes FHE the preferred choice for scenarios requiring advanced privacy-preserving computations and secure data processing.

Numerous FHE cryptosystems have been proposed, including the BFV, BGV, and CKKS schemes \cite{gorantala_cacm_fhe}. Notably, BFV and BGV schemes support integers, while the CKKS scheme extends its support to floating-point decimals. In our research, we have employed the CKKS scheme due to its compatibility with floating-point operations.

To enhance computational efficiency, we choose to pack our input into arrays of size \(2^n\) before encryption. If the input sizes are not perfect powers of 2, we pad the data with 0s. Although these ciphertexts support Single Instruction Multiple Data (SIMD) operations \cite{9481143}, they do not provide direct access to individual elements within the ciphertext.

Our research utilizes HEAAN to enable secure autonomous UAV navigation using Deep Learning. While FHE allows computations on encrypted data, certain computational operators are yet to be implemented. \textbf{In this paper, we elucidate intelligent adaptations in the encrypted domain to make encrypted data processing feasible.}


\section{Related Work}

Numerous surveys have delved into the privacy and security challenges specific to UAVs. Works such as \cite{9795697} and \cite{9488323} highlight the vulnerability landscape in UAV communication networks, emphasizing the delicate trade-off between robust security and the imperative for lightweight, efficient operations. These discussions underscore the crucial role of encryption in fortifying UAV systems against multifaceted threats, as presented by the authors in \cite{8088163}. Our research aims to build upon these foundational insights, contributing to the ongoing discourse on UAV security.

Homomorphic Encryption has been employed in prior work to secure computations in the context of UAV navigation. For instance, in \cite{alzahrani_fhe_uav}, the authors propose an extra key generation encryption technique using the Paillier Cryptosystem to secure the public key. Another approach proposes a secure and efficient method with Secure Homomorphic Encryption (SHE) for third-party UAV controllers to process client data \cite{9343124}. Additionally, Cheon et al. (Cheon et al., 2020) propose secure autonomous UAVs without relying on deep learning for autonomy, employing Linearly Homomorphic Authenticated Encryption (LinHAE).

While prior works have made significant strides in advancing autonomous systems and encryption methodologies for various applications, a comprehensive solution seamlessly integrating end-to-end encryption into a fully autonomous RL-based drone system remains elusive. Existing research has notably contributed to the field but often falls short in providing a holistic approach to ensure complete security. This study aims to bridge these gaps by introducing a novel framework that leverages RL for autonomy and integrates FHE, thereby, achieving a secure end-to-end autonomous UAV navigation system.

\begin{figure}[htbp]
\centerline{\includegraphics[scale=0.35]{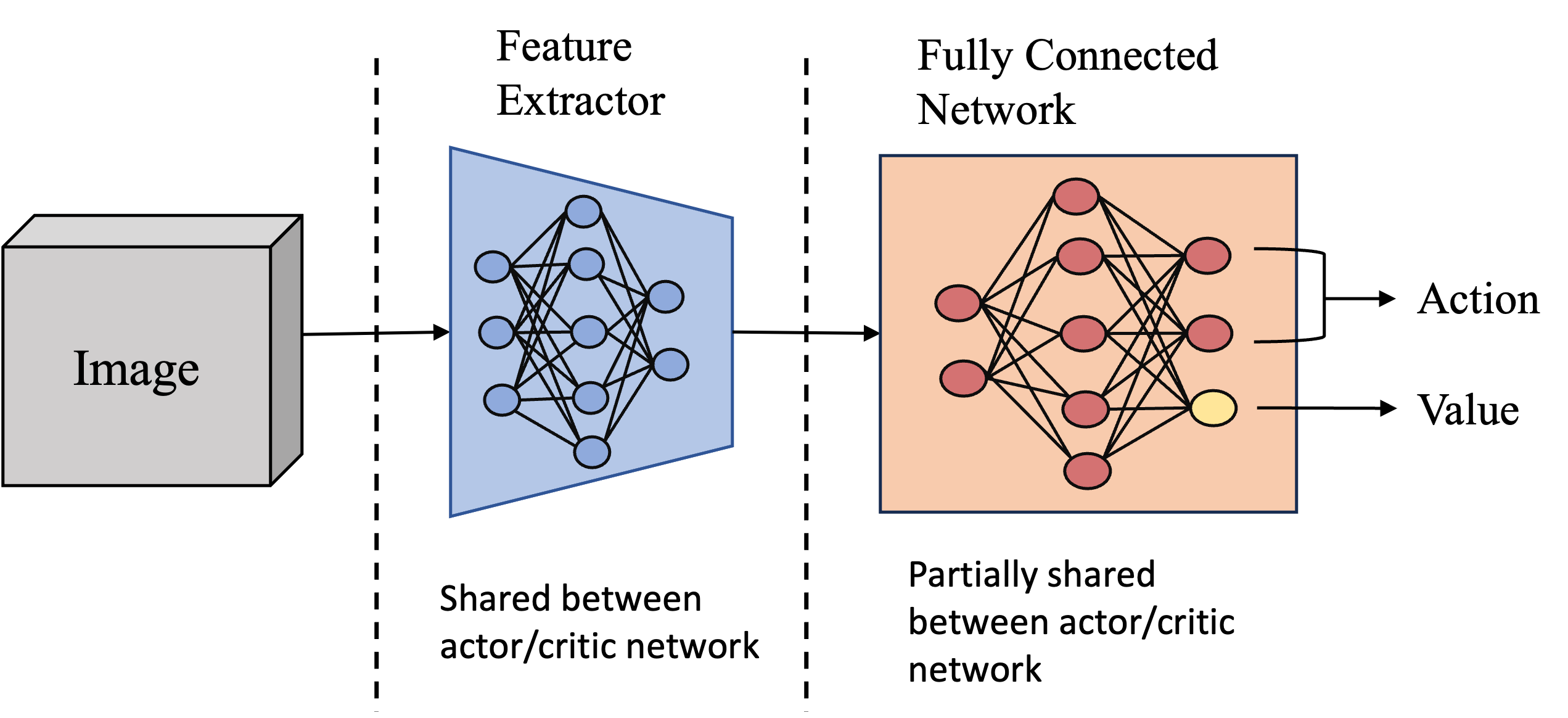}}
\caption{Architecture Overview of our Deep Learning framework implementing the Actor-Critic algorithm.}
\label{architecture_overview}
\end{figure}

\section{Proposed Method}

\begin{figure}[htbp]
\centerline{\includegraphics[scale=0.3]{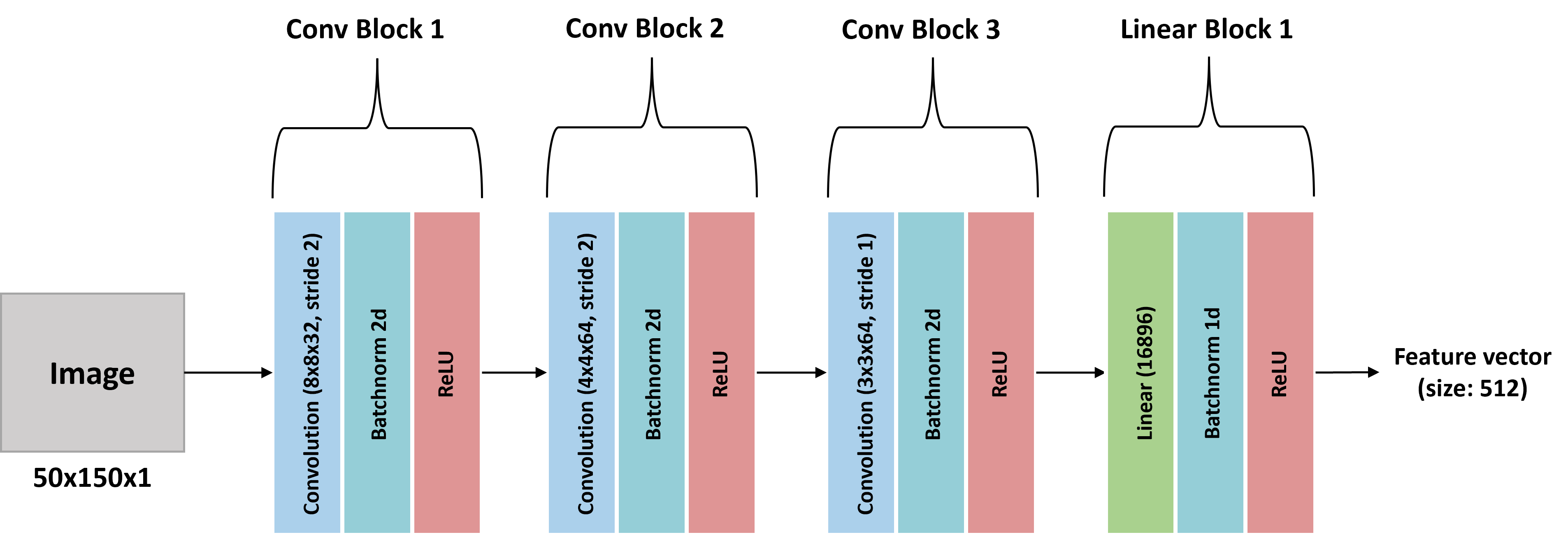}}
\caption{Feature Extractor segment of our Deep Learning framework.}
\label{feature_extractor}
\end{figure}

\begin{figure}[htbp]
\centerline{\includegraphics[scale=0.35]{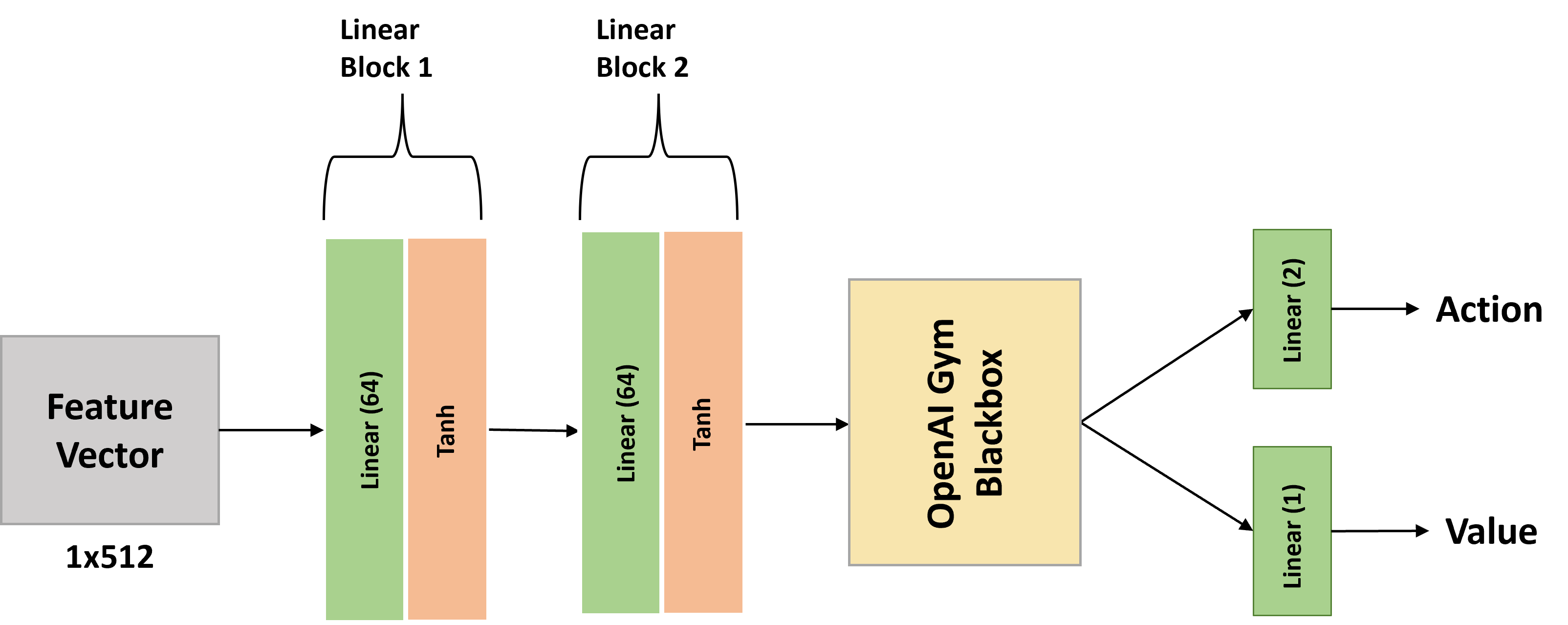}}
\caption{Fully Connected Network segment of our Deep Learning framework.}
\label{fcnn}
\end{figure}

The drone is trained using the Actor-Critic Proximal Policy Optimization (PPO) RL algorithm \cite{schulman2017ppo}. During training, both Actor and Critic networks are utilized, whereas, only the Actor network is leveraged during inferencing. The model architecture can be divided into two segments - Feature Extractor and Fully Connected Network as shown in Fig. \ref{architecture_overview}. The Feature Extractor consists of 3 convolution blocks and 1 linear block as shown in Fig. \ref{feature_extractor}. The Fully Connected Network segment consists of 2 shared linear blocks (shared between Actor and Critic) and an output linear block as in Fig. \ref{fcnn}.


Computations within the Fully Homomorphic Encryption (FHE) domain introduce several limitations - the absence of individual element access in encrypted arrays, high latency, and the absence of inherent support for operators like comparators. Consequently, we choose to train the Actor-Critic model in the unencrypted domain with data generated in a simulated environment, employing Microsoft's AirSim library and Unreal Engine. Subsequently, we leverage the model weights for inference within the encrypted domain. To achieve this, we carefully adapt each component of the RL framework to seamlessly operate within the FHE domain, addressing specific challenges presented by FHE.

The following components have been adapted to the FHE domain: (i) Input images; (ii) 2D convolution; (iii) Flattening layer; (iv) Fully connected network; (v) ReLU activation function; (iv) Tanh activation function; and (v) OpenAI Gym Library.

\subsection{Input images}

The drone's input comprises of three consecutive images, each captured from the AirSim simulator, with dimensions 50x50 (based on the simulated environment configuration). These images are concatenated to form a single input image with dimensions 50x150. HEAAN \cite{HEAAN} exclusively supports the encryption of array with sizes as powers of 2. We address this constraint by padding each row of the image with zeros, extending the width to 256. Since HEAAN supports Single Instruction Multiple Data (SIMD) operations \cite{9481143}, we try to pack as many pixel values as possible in a ciphertext. The most feasible way is to pack each row of the image as a ciphertext. Thereby, we achieve a vector of ciphertexts that represents our input image in the encrypted domain.




\subsection{Convolution Layer}

Performing regular convolution in the encrypted domain is computationally inefficient. In our research, we take a frequency-domain approach for convolution leveraging the Discrete Fourier transform (DFT). The DFT of encrypted data is performed using Homomorphic Fourier transform (HFT) - inspired by Cooley-Tukey matrix factorization \cite{d3ea2d52-5ab2-3128-8b80-efb85267295d}. 

The following steps are performed to achieve 2D convolution efficiently: 
\begin{enumerate}
    \item HFT on each ciphertext (representative of each row in the image) as in \cite{8701685} 
    \item Transpose the result based on the method in \cite{zekri_transpose}
    \item Perform HFT again on the transposed ciphertexts
    \item Transpose the ciphertexts again
    \item Compute the convolution output \(y[n]\) using element-wise multiplication in the frequency domain and Inverse DFT (Inverse HFT in the encrypted domain), as expressed in Equation \ref{eq:convolution}. 
\end{enumerate}   
\(\mathcal{G}^{-1}\) denotes the Inverse 2D DFT, and \(H[u, v]\) and \(F[u, v]\) are the 2D DFT of the ciphertext and filter, respectively.

\begin{equation}
\label{eq:convolution}
y[m,n] = \mathcal{G}^{-1}\left\{ H[u,v] \cdot F[u,v] \right\}
\end{equation}

\begin{equation}
\label{eq:2d_dft}
H[u,v] = \sum_{m=0}^{M-1} \sum_{n=0}^{N-1} h[m,n] \cdot e^{-j\frac{2\pi}{M}um} \cdot e^{-j\frac{2\pi}{N}vn}
\end{equation}

\begin{figure*}[htbp]
\centerline{\includegraphics[width=0.90\textwidth]{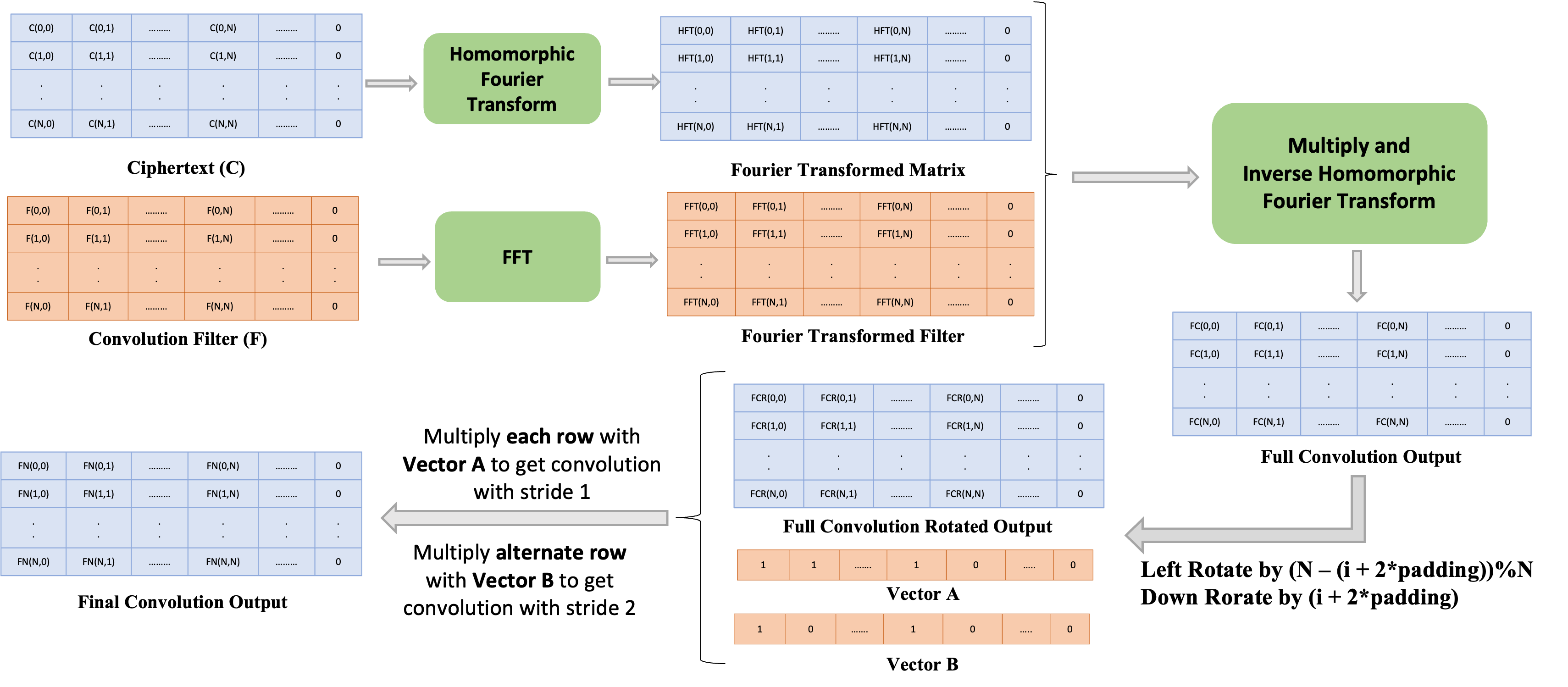}}
\caption{2D Convolution in FHE Domain. Different stride-based convolutions can be extracted by multiplying appropriate vectors. The ciphertext (representative of a row of the input image) is subjected to Homomorphic Fourier Transform (HFT) because it operates in the encrypted domain. On the other hand, the convolution filter is subjected to Fast Fourier Transform because it is in the plaintext domain.}
\label{2d_conv}
\end{figure*}

To achieve convolution with stride, a rotational manipulation is applied to the resulting ciphertext after regular convolution. We apply left rotation on the resulting ciphertext by $(N - (2 * padding)) \% N$ and down rotation by $2 * padding$, where $N$ represents the size of the ciphertext and $padding$ represents the padded value used to extract DFT convolution output. Then, this result is multiplied by an array containing 1s and 0s to obtain appropriate convolution based on the stride value, as illustrated in Fig. \ref{2d_conv}.

\subsection{Flattening layer}

The flattening operation is defined as the conversion of a 2D representation of features into a 1D representation. This is usually performed on the convolution outputs and before a dense layer. In the encrypted domain, our convolution outputs are represented as a vector of ciphertexts. Since HEAAN does not allow the concatenation of ciphertexts, the flattening operation is a computationally expensive task without decrypting and re-encrypting the ciphertexts. To perform flattening, we execute element-wise multiplication of the weights and convolution output. Element-wise multiplication is an extremely time-consuming operation as it involves multiplication, addition, and left rotation. The entire process can be summarized in 3 steps - (i) Multiply each ciphertext with its corresponding weight vector; (ii) Add it to a temporary ciphertext initialized to zeros; (iii) Perform summation on the ciphertext elements through repetitive left rotation and addition N-1 times. 

\subsection{Fully connected network}

The fully connected network is adapted to FHE as a matrix multiplication of ciphertext inputs and plaintext weight matrices. Each row of weight matrix is multiplied with the ciphertext and the elements of the ciphertext are summed through left rotation.

\begin{figure}[hbt!]
\centerline{\includegraphics[scale=0.45]{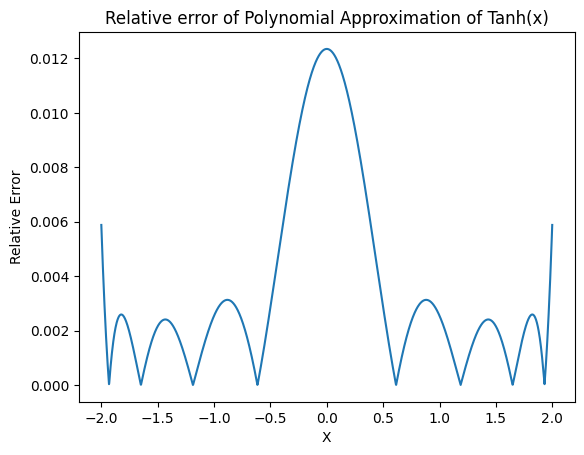}}
\caption{Relative error $\frac{\left|f(x) - 
 Tanh(x) \right|}{\left|Tanh(x)\right|}$
of f(x) over the interval [-2, 2]. f(x) corresponds to the 8-degree polynomial approximation of Tanh(x).}
\label{rel_err}
\end{figure}

\subsection{ReLU}

HEAAN lacks built-in support for comparison operations and instead uses polynomial approximations for such functions \cite{cryptoeprint:2019/417}. The CompG method is one such approximation, specifically tailored for the $-1$ to $1$ range, proposed in \cite{cryptoeprint:2019/1234} to approximate the sign function.

Normalization of input values is essential to enable the use of the CompG method to approximate ReLU. We scale down the input values to fit in the range of $[-1, 1]$ using the observed maximum absolute value during training (in the plaintext domain). The maximum value can be derived by the following formula, $\max(\left|\text{maxValue}\right|, \left|\text{minValue}\right|)$. The implementation of ReLU leverages the composite approximation technique for comparison, comparing the input value \(a\) against zero and encoding the output as 1 for \(a > 0\), 0 for \(a < 0\), and 0.5 for \(a = 0\). The ReLU output is calculated by multiplying this result by the input value \(a\). After ReLU, the values are upscaled to their original range using the inverse of the scaling factor.

\subsection{Tanh}

We use an 8-degree polynomial approximation of Tanh in FHE owing to the limitations of FHE in implementing exponential functions. As done in ReLU, normalization of inputs is an essential step before using the polynomial approximation. We restrict the input to the range $[-2, 2]$ to achieve a closer approximation. The performance of this approximation is measured through the relative error of 2000 points in the range $[-2, 2]$, as shown in Fig. ~\ref{rel_err}.

\begin{figure}[hbt!]
\centerline{\includegraphics[scale=0.35]{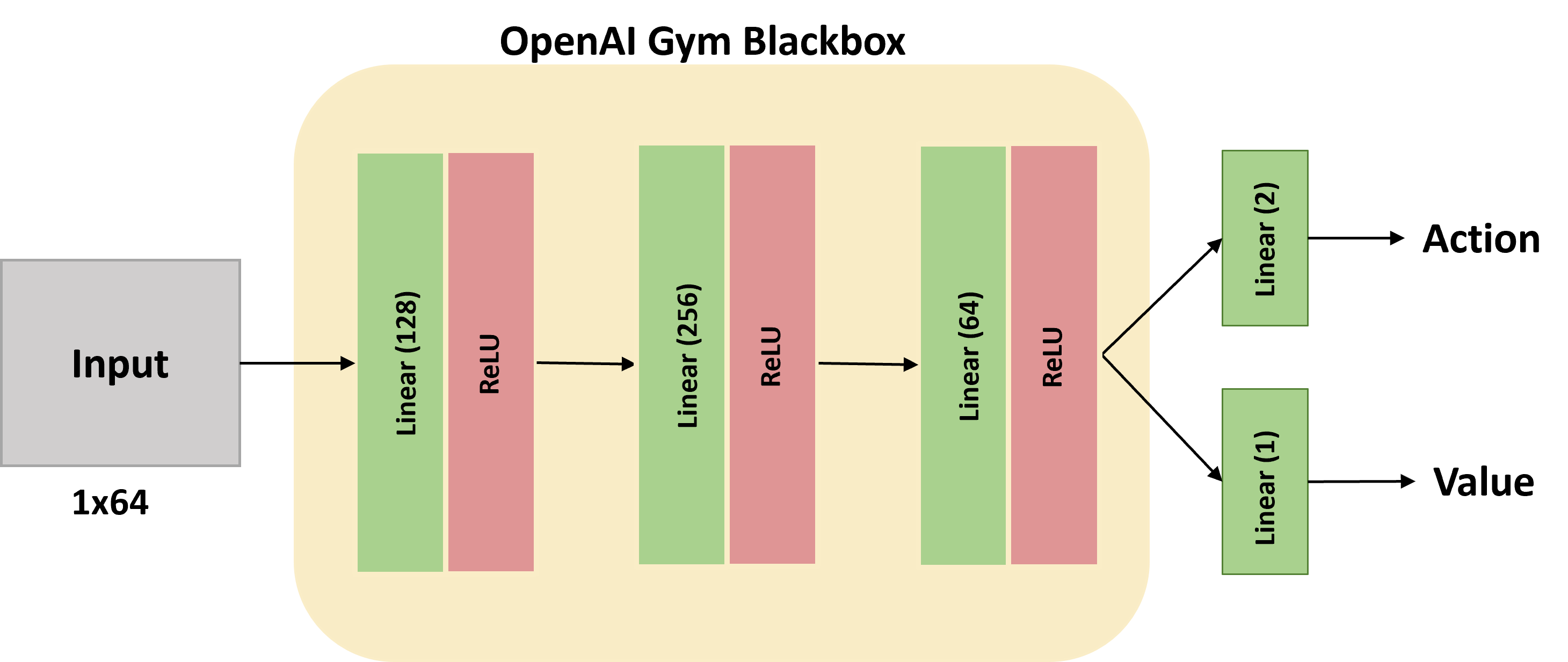}}
\caption{A neural network trained to replicate the OpenAI Gym Library to facilitate its adaptation to FHE.}
\label{gym_bb}
\end{figure}

\subsection{OpenAI Gym Library}
OpenAI Gym Library cannot be adapted directly to the encrypted domain due to the limitations of FHE in modeling probability distributions. Instead, we train a 3-layer neural network as in Fig. \ref{gym_bb} to replicate the Open AI Gym Library for our task. The neural network learns to map the final 64-dimension latent vector to the desired action output. Since the model is trained in the unencrypted domain, its weights can be used for inferencing in the FHE domain.

\section{Results}

Experiments were performed in the encrypted domain on a subset of randomly selected samples from the testing set of the plaintext domain. We compared the outcomes obtained from our FHE-enabled RL framework with those expected from the RL framework operating in the unencrypted domain. Table \ref{tab:layer_mae} depicts the mean absolute error (MAE) across each block in the network within the encrypted domain compared to its plaintext counterpart. As evident from Table \ref{tab:layer_mae}, the regression-based prediction output remained consistent between the FHE version and the plaintext counterpart for the tested samples. In addition to low MAE scores, we also achieve an \textbf{R-squared score of 0.9631} with the end-to-end FHE-based Reinforcement Learning framework, in comparison with results in the unencrypted domain. Further, Table \ref{tab:layer_time} presents the average processing time across each block in the network. These findings substantiate the efficacy of our FHE-adapted network, showcasing the viability of FHE in preserving model accuracy while ensuring data confidentiality.

\begin{figure}[hbt!]
\centerline{\includegraphics[scale=0.45]{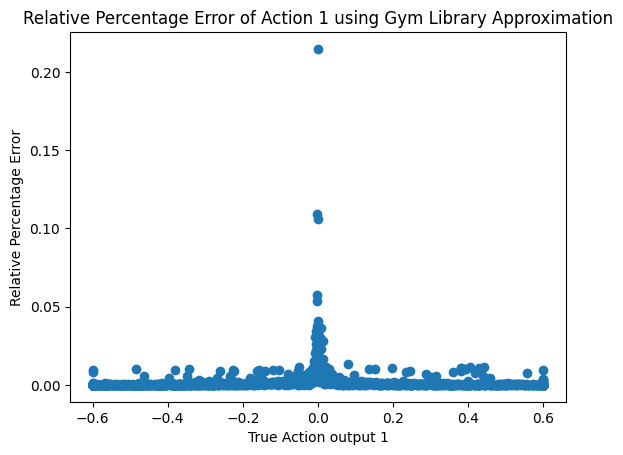}}
\centerline{\includegraphics[scale=0.45]{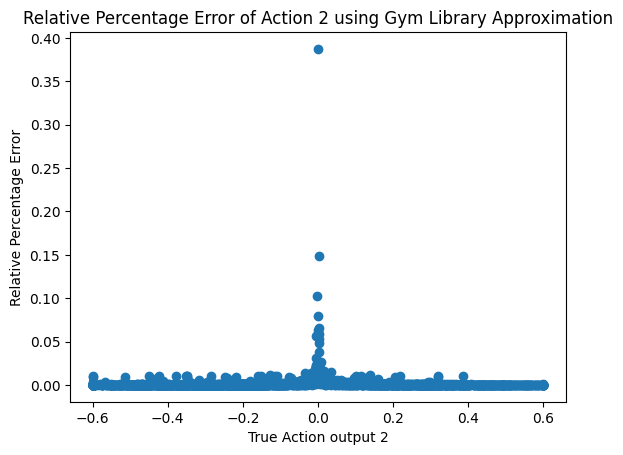}}
\caption{Relative percentage errors of actions (navigational results) after adapting the OpenAI Gym Library to FHE.}
\label{rel_err_actions}
\end{figure}


\begin{table}[ht]
\centering
\caption{Mean Absolute Error (MAE) between plain-text and FHE model intermediate outputs for each block in the network.}
\label{tab:layer_mae}

\renewcommand{\arraystretch}{1.5}

\begin{tabular}{|l|l|}
\hline
\textbf{Layer} & \textbf{MAE}         \\ \hline
Convolution Block 1  & 0.0741          \\ \hline

Convolution Block 2  &  0.0971  \\ \hline
Convolution Block 3  &  0.0626                     \\ \hline
Linear Block 1    & 0.0105   \\ \hline                    
Linear Block 2  & 0.0184                    \\ \hline
Linear Block 3  & 0.0098                   \\ \hline

OpenAI Gym Library Blackbox  & 0.0210                   \\ \hline
\end{tabular}
\end{table}

\begin{table}[!h]
\centering
\caption{Time taken by each block in the FHE-adapted network.}
\label{tab:layer_time}

\renewcommand{\arraystretch}{1.5}

\begin{tabular}{|c|c|}
\hline
\textbf{Layer}       & \textbf{Inference Time (seconds)} \\ \hline
Convolution Block 1  &  9471.27                  \\ \hline
Convolution Block 2  &  280831.67                    \\ \hline
Convolution Block 3  &  716034.24                     \\ \hline
Linear Block 1    & 12069.36 \\ \hline
Linear Block 2  &  790.5                  \\ \hline
Linear Block 3  &  802.62                  \\ \hline
OpenAI Gym Library Blackbox  & 4754.82 \\ \hline
\end{tabular}
\end{table}

\section{Conclusion}

In this paper, we propose to use a combination of FHE and deep neural networks to enable the secure navigation of autonomous UAVs. We show that navigation results from communication networks can be strictly secure without a loss in performance. We demonstrate our approach by adopting an RL framework for autonomy. Since FHE cannot perform all mathematical operations, we detail the adaptation of each layer of the RL framework to FHE - input images, convolutional layers, fully connected networks, activation functions, and the OpenAI Gym Library. Evaluations of the proposed framework demonstrate minimal mean absolute error (MAE) across each block in the network and a high R-squared score, showcasing no discernible accuracy loss when compared to its plaintext counterpart. Since FHE guarantees are based on strong theoretical principles, privacy and security are ensured, and only authorized individuals with the secret key will be able to access the results from the FHE computation. Future efforts will be directed towards substantially decreasing inference time.


{\small
\bibliographystyle{unsrt}
\bibliography{uavNav}
}

\end{document}